\RequirePackage{fix-cm}
\documentclass[twocolumn]{svjour3}          % twocolumn
\smartqed  % flush right qed marks, e.g. at end of proof
\usepackage{mathptmx}
\usepackage{amstext}
\usepackage{graphicx}

\journalname{Granular Matter}

\begin{document}

\title{Particle characterization using THz spectroscopy}

\author{Philip Born         \and
        Karsten Holldack		\and
        Matthias Sperl
}

\institute{P. Born, M. Sperl \at
              Institut f\"ur Materialphysik im Weltraum, Deutsches Zentrum f\"ur Luft- und Raumfahrt, 51170 Cologne, Germany  \\
              Tel.: +49-2203-601-4048\\
              Fax: +49-2203-61768\\
              \email{philip.born@dlr.de}           %  \\
           \and
           K. Holldack \at
              Helmholtz-Zentrum Berlin f\"ur Materialien und Energie GmbH, Albert-Einstein-Str. 15, 12489 Berlin, Germany
}

\date{Received: date / Accepted: date}

\maketitle

%--------------------------------------------------------------------------------------------------

\begin{abstract}
THz extinction spectroscopy extends UV-Vis and NIR-spec\-tros\-copy to characterize particles from fine powders and dust to sand, grains and granulated materials. We extract particle sizes from the spectral position of the first peak of the interference structure and size distributions from the visibility of the fine ripple structure in the measured extinction spectra. As such, we can demonstrate a route for a quick determination of these parameters from single measurements.
\end{abstract}

%--------------------------------------------------------------------------------------------------

\section{Introduction}
\label{intro}

Currently emerging THz technologies offer new ways for investigations of granular particle ensembles.  The wavelength of THz radiation in the range of 30~$\mu$m to 3~mm \cite{Bundermann} matches particle sizes in many typical granular media. Consequently, situations known from measurements in the colloidal regime using visible light or from x-ray based measurements on atomic and molecular structures can be reproduced. Many quantities obtainable from angle-resolved scattering and spectrally resolved transmission measurements become highly sensitive to particle size and other geometric features of the samples, such as degree of polarization, scattered intensity and extinction \cite{Hulst}. 

Light scattering methods for granular media like laser diffraction, presently using visible light, offer significant advantages over image analysis and analytical sieving, such as fast and potentially in-situ screening of large particle numbers \cite{Eshel2004,Kelly2006,Blott2006}. Angle-resolved static light scattering measurements in the THz regime from granular media have been shown to deliver particle sizes and packing characteristics of thin samples \cite{Born2014}.

Here we show the possibility to characterize particles in granular media using broad-band THz spectroscopy, a method that offers some advantages compared to monochromatic angle-resolved measurements. Spectroscopic measurements can be performed both in transmission or reflection, without the need to move the detector during the measurement, and with multiple-scattering samples where only little information is gained from angle-resolved scattering. Thus, even samples with limited accessibility can be remotely characterized. Spectroscopic measurements in the ultraviolet, visible and infrared spectral ranges are a common approach in remote sensing to determine particle sizes \cite{King1978,Clancy2003,Korablev1993,Mathis1977} and are applicable for process monitoring in industrial processing of powders \cite{Higgins2003,Pasikatan2001}. Extending the spectral range to the THz region allows for extending these remote measurement techniques to the domain of sands, soils, grains and granulated powders. 

We present experimental THz extinction spectra of granular media which illustrate the sensitivity to the size and the size distribution of macroscopic particles. We reconstruct the extinction spectra by taking into account the detector resolution and the particle size distribution. Relations are inferred from these calculations to derive the particle size distribution from measured extinction spectra. The results are verified using light microscopy and image analysis. Finally, we discuss the applicability and limitations of the method.

%--------------------------------------------------------------------------------------------------

\section{Experimental}
\label{sec:experiment}

    \begin{figure}
		\centering
  	\includegraphics[width=0.4\textwidth]{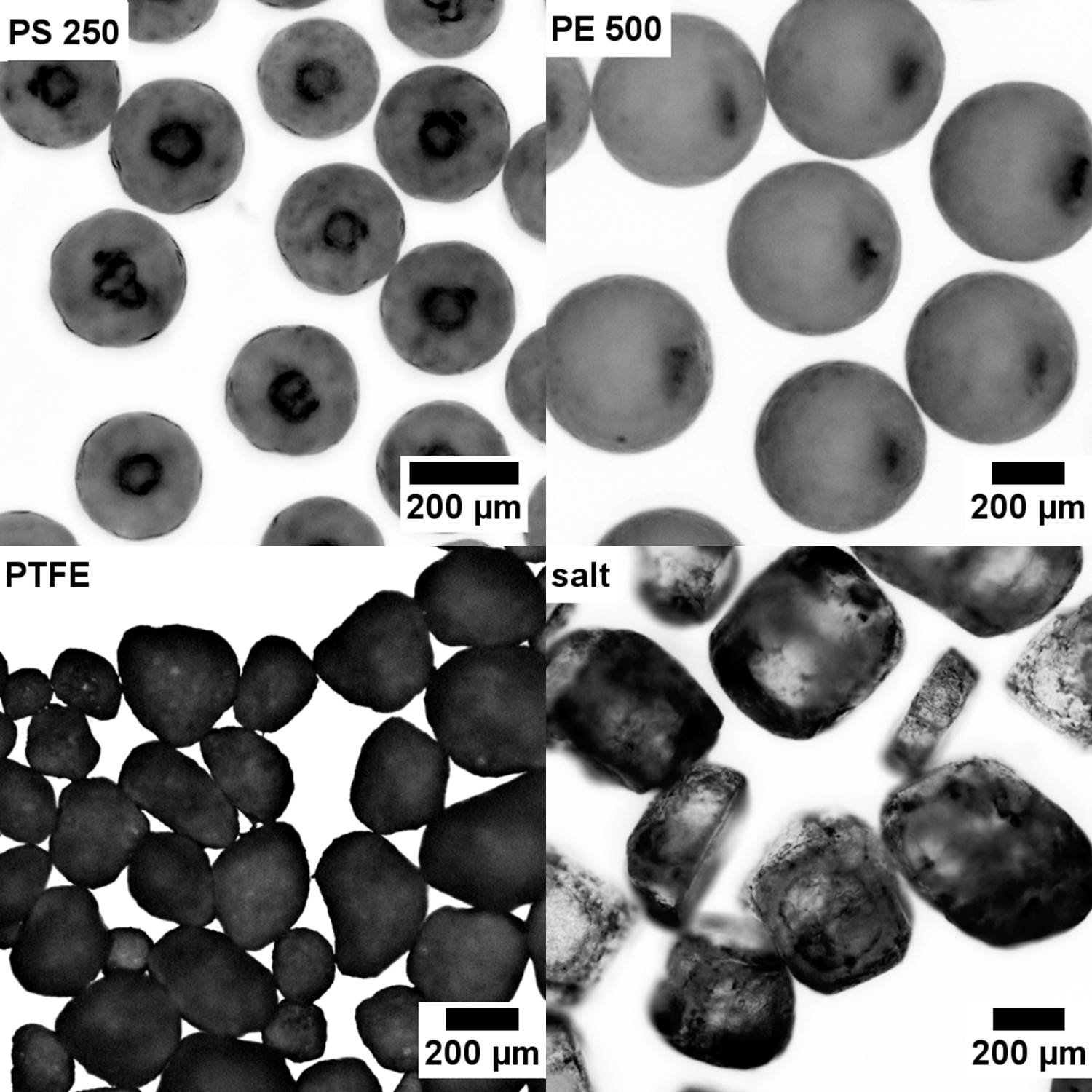}
		\caption{Light microscopy images of the nearly monodisperse spherical polystyrene (PS, here PS 250) and polyethylene (PE, here PE 500) samples and the technical grade Teflon (PTFE) and salt samples used in this work.}
		\label{fig:5}       % Give a unique label
	\end{figure}
The measurements were carried out in transmission geometry under vacuum (0.1~mbar) using the high-resolution FT-IR spectrometer Bruker IFS 125 HR available at the BESSY II storage ring \cite{Holldack2007}. The spectral range of the measurements shown here covered wavenumbers $\nu$~= 20 \ldots 370~cm$^{-1}$ (i.e. wavelengths $\lambda$~= 500 \ldots 27~$\mu$m, or frequencies $f$ = 0.6 \linebreak\ldots 11.1~THz). In this particular case the spectral range was limited by the selected source (internal Hg-lamp), the 6~$\mu$m multilayer-mylar beamsplitter and the detector, a 4.2~K Si-Bolometer (Infrared Labs). Using coherent synchrotron radiation, a 1.6~K Si-Bolometer and a 125~$\mu$m beamsplitter transmission measurements down to 2~cm$^{-1}$ are feasible \cite{Abo2003}. The resolution was selected at 0.5 cm$^{-1}$ (maximum resolution of the instrument is 0.0063~cm$^{-1}$). Two iris diaphragms in front and behind the sample holder with inner diameters of 10~mm constrained and collimated the radiation.

Particles were either fixed in the light path on transparent adhesive tape or within thin sample compartments made from polyethylene foil. Samples were nearly monodisperse spherical particles made from polystyrene (PS, Microbeads Dynoseeds, nominal diameter 80, 140, 250 and 500~$\mu$m) or polyethylene (PE, Cospheric CPMS, nominal diameter 90, 250 and 500~$\mu$m), and technical grade Teflon (PTFE, 3M Dyneon PTFE Granules TF1641), salt and sugar (Fig.~\ref{fig:5}). The extinction spectra were obtained by measuring the extinction of the empty sample holder as reference and subsequently the sample holder with particles. In case of measurements with sample compartments the PE foil windows were gently inclined in order to suppress Fabry-P\'erot oscillations in the reference spectra. 

The extinction $E$ of the samples was obtained from the measured intensity with sample $I_{1}$ and the reference intensity $I_{0}$ by $E = - ln \left ( \frac{I_{1}}{I_{0}} \right )$. Lambert-Beer's law suggests that this extinction is proportional to the extinction efficiency $Q_{e}$ of the particles up to a proportionality factor,
	\begin{equation}
		I_{1} = I_{0} \cdot e^{-\Phi \cdot A_{p} \cdot L \cdot Q_{e}} \equiv I_{0} \cdot e^{-E},
		\label{eq:lbl}
	\end{equation}
with $\Phi$ the number density of particles, $A_{p}$ the geometrical cross-sectional area of the particles, and $L$ the sample thickness in direction of the beam. We scale all extinction spectra to the same maximum value to compare to theoretical predictions.

The spectroscopy results on particle size and size distribution were compared to results from image analysis for verification. Particles were fixed on carbon tape and investigated in a Zeiss AxioImager.A1m microscope in reflection mode. Roughly 100 to 600 particles are analyzed using the ImageJ software \cite{imagej} to obtain the particle radii and the radius distributions.

The measured extinction spectra are compared to theoretical predictions of the spectral extinction efficiency by the Mie theory for scattering from monodisperse spheres. We used refractive indices of PS ($m\approx1.59+i0.002$) and PE ($m\approx1.55+i0.001$) available from the literature \cite{Cunningham2011} and the radii of the particles from light microscopy as input parameters. The actual calculations of the Mie extinction efficiencies were performed using a MatLab-implementation of the code proposed by Boh\-ren and Huff\-man \cite{BH1983,Matzler2002}.

%--------------------------------------------------------------------------------------------------

\section{Results}
\label{sec:results}
	
	\begin{figure}
  	\includegraphics[width=0.4\textwidth]{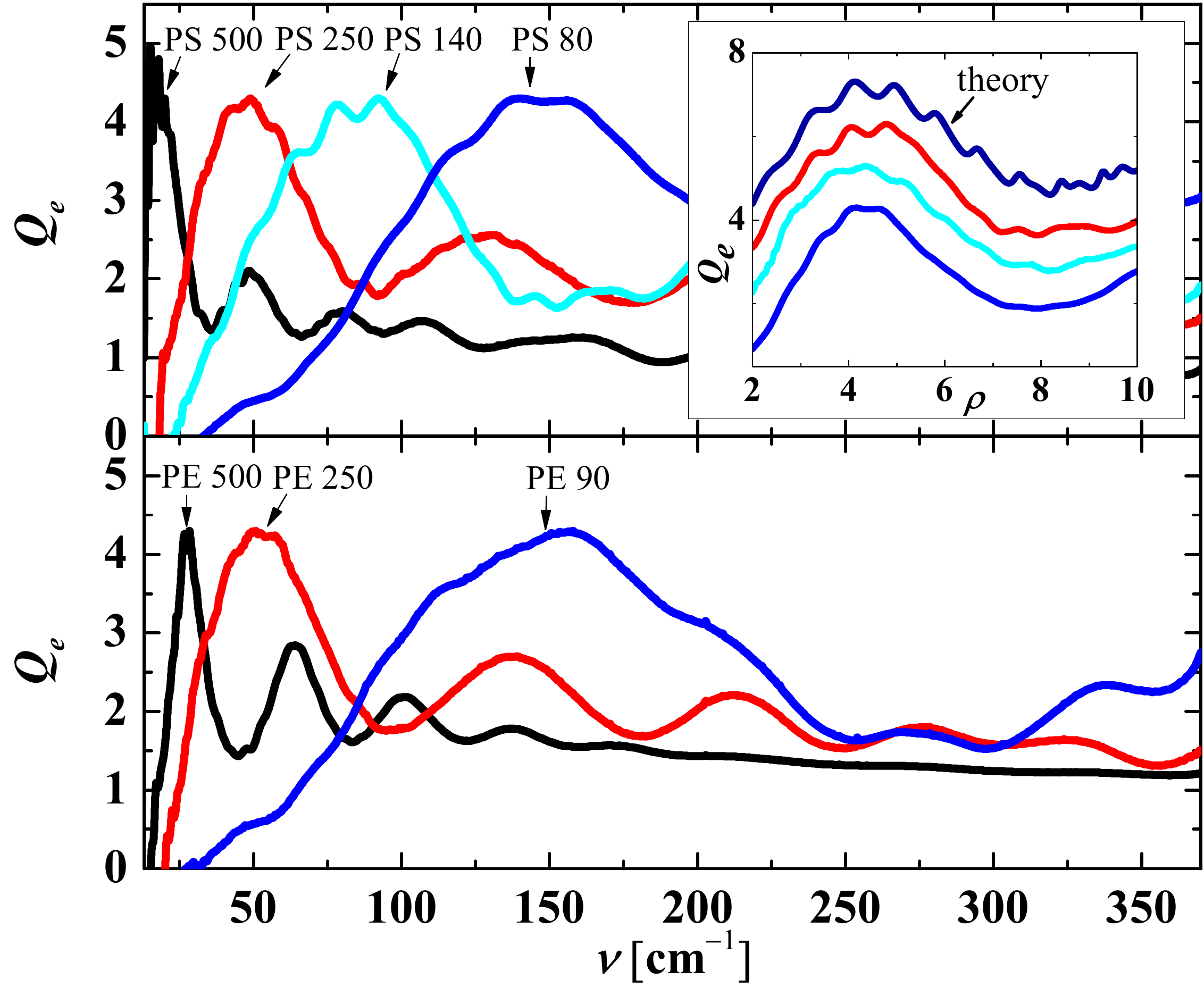}
		\caption{Measured extinction efficiencies $Q_{e}$ of PS (upper panel) and PE (lower panel) particles. The labels indicate nominal particle diameters in $\mu$m. The inset in the upper panel shows a close-up of the first extinction peak plotted as a function of the phase shift $\rho$ created by the PS particles. The offset curves are sorted by the polydispersity of the particles (color-coding as in the main diagram). The top curve is the calculated extinction efficiency of monodisperse particles.}
		\label{fig:1}       % Give a unique label
	\end{figure}
Figure~\ref{fig:1} depicts the measured spectroscopic extinction for the PS and PE particles. The spectra all show a characteristic strong extinction peak with successive decaying oscillations. The extinction maximum shifts towards smaller wavenumbers $\nu$ with increasing particle size. A quantitative relation among the extinction spectra and the particle size can be inferred from the phase shift $\rho = 2\cdot x\cdot |m-1|$ created by the particles, where $x = 2\pi a/\lambda = 2\pi a\cdot \nu / 10000 $ is the size parameter of a particle with radius $a$ at a wavelength $\lambda$ and $m$ is the refractive index of the particles. In particular, the first extinction maximum occurs at a fixed phase shift created by the particles of $\rho \approx 4.2$ \cite{Hulst}. The measured extinction spectra consequently collapse on a single curve when plotted as a function of the phase shift (see inset in Fig.~\ref{fig:1}). Picking the wavenumber $\nu_{max}$ at the first extinction maximum gives a quick estimate of the mean particle size,
	\begin{equation}
		a_{\text{spec}} \approx \frac{4.2 \cdot 10000}{4\pi |m-1|\cdot \nu_{max}}
		\label{eq:a}
	\end{equation}

The maximum position from the measurements and relation~(\ref{eq:a}) together with refractive indices available from the literature \cite{Cunningham2011} gave the particle radius $a_{\text{spec}}$. A good agreement with particle radii from image analysis is found (see below, Fig.~\ref{fig:4}). 

    \begin{figure}
  	\includegraphics[width=0.4\textwidth]{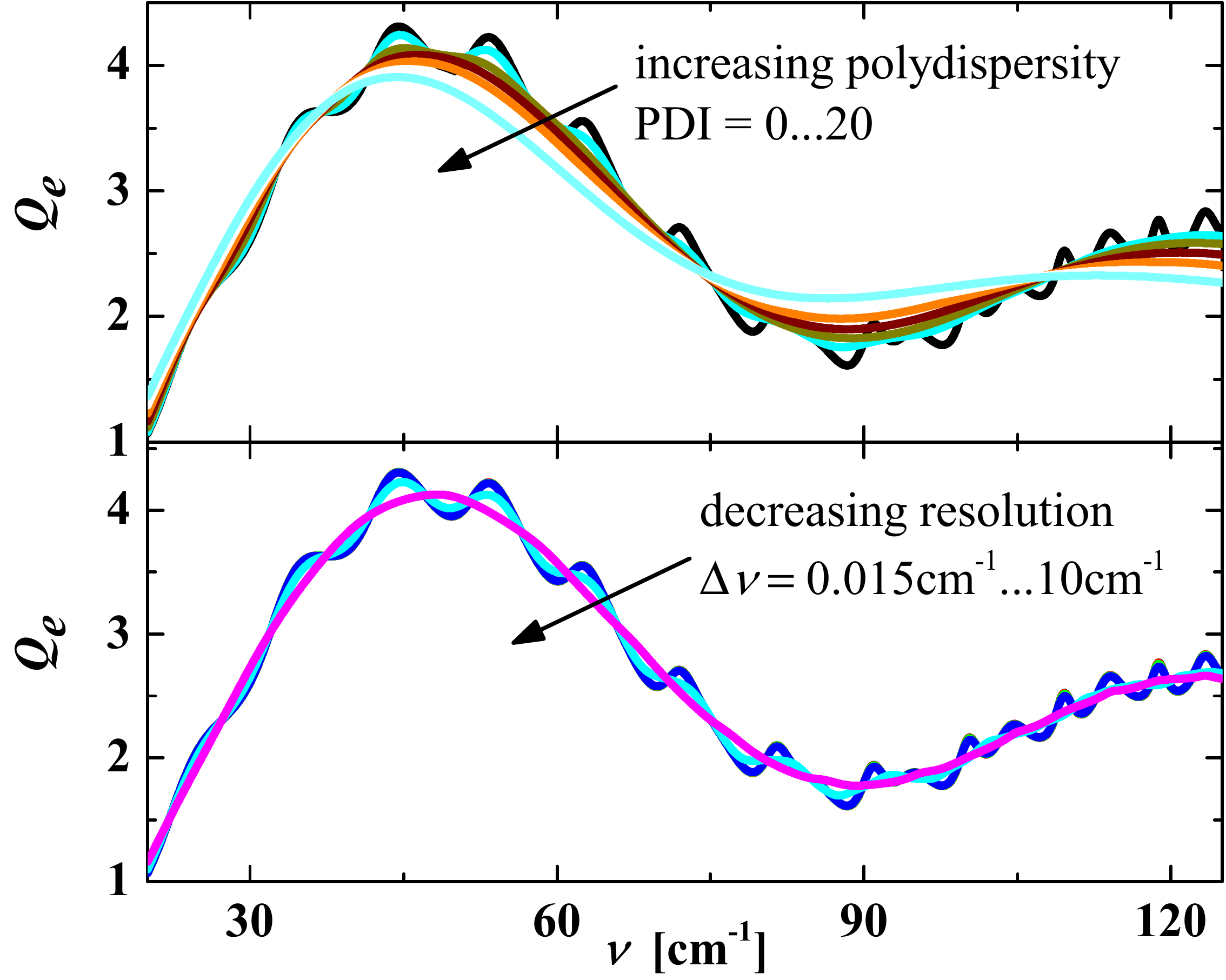}
		\caption{Calculated extinction efficiencies $Q_{e}$ of particles with radius $a=125\mu$m and the refractive index $m_{PS}$ of polystyrene. The upper diagram shows the effect of increasing the polydispersity index PDI of the particles from 0 to 20 (in steps of 3), which first smears out the ripple structure and consequently the interference structure. The lower diagram illustrates the influence of detector resolutions $\Delta\nu$ of 0.015~cm$^{-1}$, 0.1~cm$^{-1}$, 0.5~cm$^{-1}$, 1~cm$^{-1}$, 5~cm$^{-1}$, and 10~cm$^{-1}$. The decreasing resolutions suppress the visibility of the ripple structure.}
		\label{fig:2}       % Give a unique label
	\end{figure}
The measured extinction, however, differs from the prediction for the extinction efficiencies of monodisperse spheres as obtained from Mie theory (see inset in Fig.~\ref{fig:1}). Extinction by dielectric spheres features a series of maxima and minimima with a strong leading maximum, qualitatively explainable by favorable and unfavorable interference of light diffracted and transmitted by the particles, and a finer superimposed ripple structure \cite{Hulst}. Two important aspects limiting the visibility of those features are the selected spectral resolution of the instrument and the size distribution of the particles in the sample under investigation. We take advantage of these dependencies in order to retrieve the particle size distribution, or polydispersity, from the measurements.

    \begin{figure}
  	\includegraphics[width=0.4\textwidth]{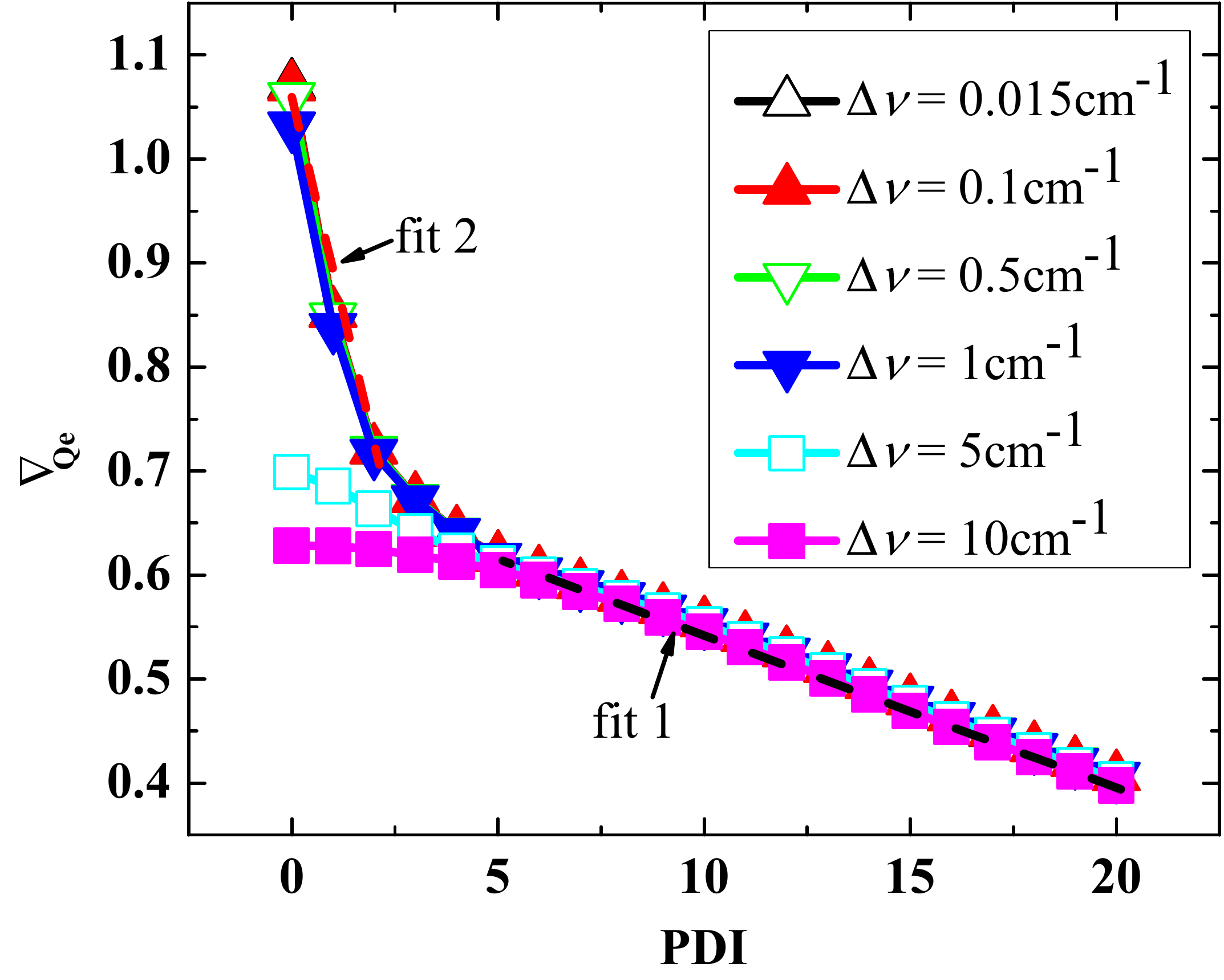}
		\caption{The relation between the polydispersity and the average slope $\nabla_{Q_{e}}$ of the calculated extinction spectra. The calculated average slope $\nabla_{Q_{e}}$ of the respective extinction spectrum is plotted as a function of the polydispersity index PDI used in calculating $Q_{e}$, for different detector resolutions $\Delta\nu$. Two linear relations for the PDI as a function of the average slope are found by fitting, depending on the detector resolution.}
		\label{fig:3}       % Give a unique label
	\end{figure}
The measured extinction efficeincy $Q_{e}$ at each data point $\nu_{i}$ is actually a smoothed value given by the spectral resolution of the detector:
	\begin{equation}
		Q_{e}(\nu_{i}, \Delta\nu) = \frac{1}{\Delta\nu} \int_{\nu_{i}-(\Delta\nu/2)}^{\nu_{i}+(\Delta\nu/2)} Q_{e}(\nu)d\nu,
	\label{eq:resolution}
	\end{equation}
where $\Delta\nu = |\nu_{i} - \nu_{i+1}|$ defines the spectral resolution of the detector. Figure~\ref{fig:2} shows the smoothing effect of several detector resolutions $\Delta\nu$. This smoothing is overlayed to the effect of polydispersity in a measurement and may limit the sensitivity to the particle size distribution.

The polydispersity of the particles induces an effective extinction efficiency which depends on the particle radii distribution $N(a)$ \cite{Hulst},
	\begin{equation}
 		Q_{e,eff}(\nu_{i}, N(a)) = \frac{\int_{0}^{\infty}{Q_{e}(\nu_{i},a)\cdot \pi a^{2}\cdot N(a)da}}{\int_0^{\infty} \pi a^{2} \cdot N(a) da}. 
	\label{eq:qepoly1}
	\end{equation}
We suggest to avoid the costly and time-consuming computation of $Q_{e}$ for many radii $a$ to perform the integral, as $Q_{e}$ only depends on the phase shift $\varrho$. It is sufficient to calculate $Q_{e}$ for one $a_{0}$ and then shift the $Q_{e}$ spectrum for each $a$ according to 
	\begin{equation}
 		Q_{e,eff}(\nu_{i}, N(a)) = \frac{\int_{0}^{\infty}{Q_{e}(\nu=\frac{a}{a_{0}}\cdot\nu_{i},a_{0})\cdot \pi a^{2}\cdot N(a)da}}{\int_0^{\infty} \pi a^{2} \cdot N(a) da} 
	\label{eq:qepoly2}
	\end{equation}
We assume a Gaussian distribution of the particle radii $N(a)$ around a central radius $a_{0}$ to investigate the influence of the polydispersity on the extinction. The relative width of the size distribution is conveniently described by the polydispersity index PDI derived from the the standard deviation $\sigma$, $\text{PDI} = (\sigma/a_{0})\cdot 100 $. Figure~\ref{fig:2} shows the calculated effective extinction efficiencies. The finer structures of the extinction spectra vanish and the interference peaks shift slightly toward smaller wavenumbers with increasing polydispersity.

The smoothing of the extinction spectrum by the polydispersity suggests a useful measure for the polydispersity. We characterize the spectra given in Fig.~\ref{fig:2} by the average absolute slope of the curves $\nabla_{Q_{e}}$, 
    \begin{equation}
        \nabla_{Q_{e}}\equiv\overline{\left|\frac{dQ_{e}}{d\rho}\right|}\approx \overline{\left| \frac{Q_{e}(\nu_{i})-Q_{e}(\nu_{i-1})}{\rho_{i}-\rho_{i-1}}\right|},
    \end{equation}
which we calculate for $\rho$ from 2 to 10 in order to compare with the experiments. Figure~\ref{fig:3} plots the calculated slope of the extinction as a function of the polydispersity index PDI for different detector resolutions $\Delta\nu$. The maximum detectable slope of the extinction is set by the detector resolution, and increasing polydispersity continuously lowers the slope. We fit two straight lines to derive functions which allow us to calculate the polydispersity from a measured slope of the extinction. For slopes below 0.6 we obtain (fit 1):
	\begin{equation}
		\text{PDI} = 47.9 - 68.7\cdot \nabla_{Q_{e}}.
	\label{eq:pdi1}
	\end{equation}
Higher slopes are only well resolved by measurements with resolutions $\Delta\nu\leq1$~cm$^{-1}$. In this regime, with slopes larger than 0.7, we obtain (fit 2):
	\begin{equation}
		\text{PDI} = 5.9 - 5.6\cdot \nabla_{Q_{e}}.
	\label{eq:pdi2}
	\end{equation}

	\begin{figure}
	  \includegraphics[width=0.4\textwidth]{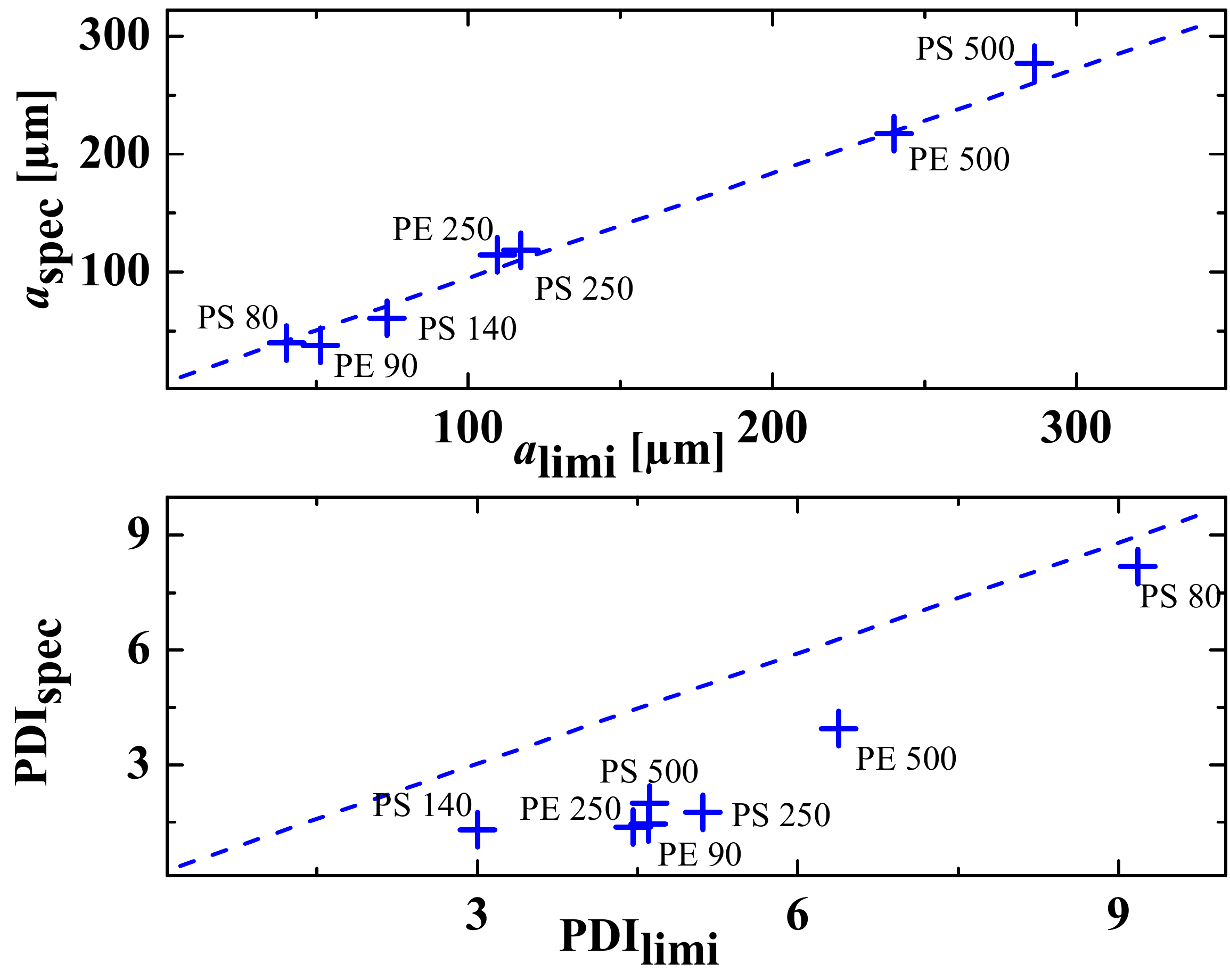}
		\caption{Results of the particle characterization using THz spectroscopy. The upper diagram shows the particle radii measured using THz spectroscopy ($a_{\text{spec}}$) plotted versus the radii measured using light microscopy ($a_{\text{limi}}$). The lower diagram shows the polydisersity index measured using THz spectroscopy (PDI$_{\text{spec}}$) plotted versus the polydispersity measured using light microscopy (PDI$_{\text{limi}}$). The dashed lines give the diagonals, $a_{\text{spec}} = a_{\text{limi}}$ and PDI$_{\text{spec}}$~=~PDI$_{\text{limi}}$ }
		\label{fig:4}       % Give a unique label
	\end{figure}
With the particle radius $a_{\text{spec}}$ known from relation (\ref{eq:a}) the extinction spectra can be plotted versus the phase shift $\varrho$ (inset in Fig.~\ref{fig:1}). With a common abscissa the average slope of the measured extinction $\nabla_{Q_{e}}$ can be calculated and evaluated using the relations (\ref{eq:pdi1}) and (\ref{eq:pdi2}). The obtained PDIs can again be compared with PDIs determined  by visible light image analysis. Figure~\ref{fig:4} depicts the results of this evaluation. 

	\begin{figure}
  	\includegraphics[width=0.4\textwidth]{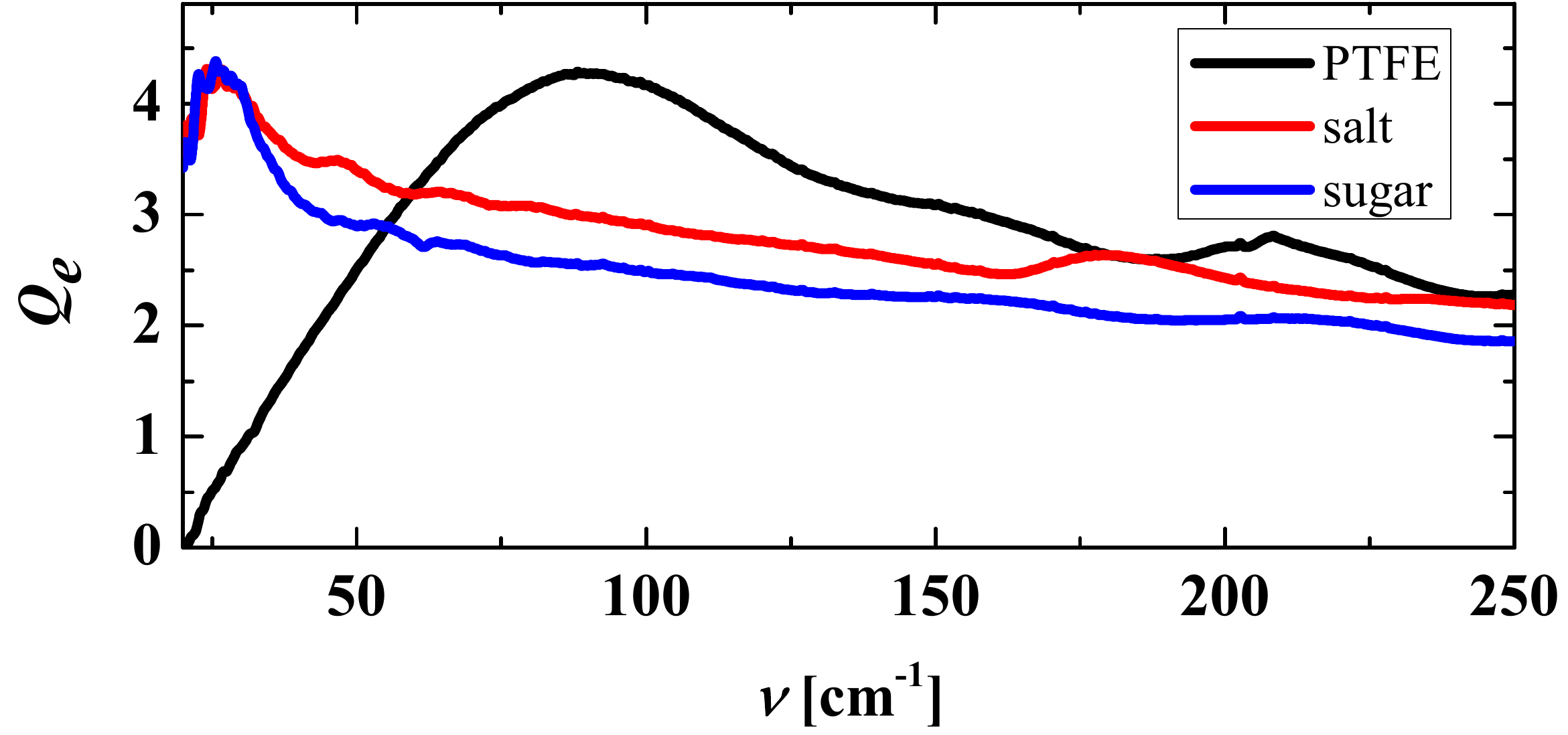}
		\caption{Measured THz extinction efficiencies of polydisperse, nonspherical particles. PTFE, salt and sugar particles exhibit extinction peaks which allow to extract average equivalent sphere radii.}
		\label{fig:6}       % Give a unique label
	\end{figure}
Finally, the applicability to technical grade samples was tested. Without further treatment PTFE, salt and sugar were spread on adhesive tape and measured. In Figure~\ref{fig:5} light microscopy images of the samples are displayed (sugar gave similar images as salt). The particles were non-spherical and exhibited broad size variations. Determination of the particle size in such samples using image analysis would be laborious, if not impossible, as the broad size variation demands high particle numbers to be evaluated, while the irregularity makes automated detection difficult. The THz extinction spectra of these samples (Fig.~\ref{fig:6}) reflect these properties, albeit extraction of averaged information on these measurements is much faster and more direct than from image analysis. The extinction spectra of PTFE, salt and sugar exhibit distinct first extinction maxima, which immediately allow to extract average equivalent sphere radii ($a_{\text{salt, sugar}}\approx 220~\mu m, a_{\text{PTFE}}\approx 65~\mu m$, using an assumed $m=1.6$).

%--------------------------------------------------------------------------------------------------

\section{Discussion}

The results presented in Fig.~\ref{fig:4} demonstrate that THz extinction spectroscopy is a promising method for measuring particle sizes and size distributions in granular media. The particle radii determined from spectroscopy reproduce the radii from light microscopy very well, with a correlation coefficient of 0.995. The size distributions determined by the two methods clearly follow the same trend (with a correlation of 0.96), however, the spectroscopic PDI are offset towards smaller values compared to the image analysis results. The reason is not clear yet. Potential factors influencing the measurement results have to be investigated in detail, such as sample preparation, image processing and particle number statistics (only 100-600 particles are investigated with image analysis) and the assumed ideal Gaussian distribution for deriving relations (\ref{eq:pdi1}) and (\ref{eq:pdi2}).

The potential applicability of the method motivates a discussion of the limitations. The precision of the measurements is influenced by three factors: a good estimate of the refractive index $m$ in the spectral region under consideration, a well resolved first extinction peak of the interference structure, and an extinction that is caused by undisturbed scattering by individual particles. 

The refractive index $m$ in the THz regime is already available for many materials from the literature \cite{Cunningham2011,Piesiewicz2007,Birch1992,Folks2007}. It can also be measured using THz spectroscopy and bulk samples \cite{Scheller2009}. However, while many unpolar and disordered materials like PE and PS have nearly constant refractive indices in the THz range, other materials may exhibit strong variations in the refractive index which cause spectroscopic features superimposing the interference structure. In such cases, the extinction spectra must be evaluated with care.

A well resolved extinction can only be obtained if the spectral range is adapted to the particle size. This defines the measurement range of the method. In the experiments presented here we used a spectral range of $\nu$~= 20~\ldots 370~cm$^{-1}$, which limits the resolvable particle radii to $a\approx$ 300~\ldots 15~$\mu$m (using eq.~\ref{eq:a}). Using other instrument settings and coherent synchrotron radiation from the storage ring BESSY II, measurements down to 2~cm$^{-1}$ can be performed and consequently measurement ranges up to particle radii $a\approx$ 5~mm are feasible. These lower wavelength limits are mainly given by the quasi-optical properties of the setup. They come into reach even with modern THZ-TDS spectrometers \cite{Bundermann}, but we note here that an in-vacuum operation of these experiments is mandatory to avoid spectral distortions by atmospheric extinction.

The sensitivity of spectroscopy to the particle size originates from the extinction of light passing through the sample by scattering \cite{BH1983}. The method thus can only be expected to deliver reliable results when the scattering process of the particles is undisturbed. Disturbances of the scattering process happen when the particles in the sample come too close together. The scattering process becomes disturbed in this case by positional correlations among the sample and interference of light scattered from adjacent particles \cite{Fraden1990} and by coupling of the scattering of particles within the near-field of each other \cite{Mishchenko1990}. Although this limits the applicability of spectroscopy to determine particle in terms of packing density, it goes far beyond laser diffraction with visible light, which is restricted to single scattering samples \cite{Eshel2004}. Detrimental effects of the packing density on the extinction spectrum were estimated to appear above packing fractions of 0.2 \cite{Quirantes2001}.

%--------------------------------------------------------------------------------------------------

\section{Conclusions}

We demonstrated a particle characterization method in granular media using THz extinction spectroscopy. Particle sizes and size distributions were obtained that reproduced results obtained using visible light image analysis. With illuminated areas in the square centimeter range and potential application to multiple scattering samples, thousands of particles are investigated simultaneously, giving the measurements a high validity. These benefits come at the cost of dependence on more parameters than image analysis or laser diffraction within the Fraunhofer approximation. Extinction depends on the particle size and the complex refractive index of the particles. Exact measurements consequently require knowledge of the refractive index of the material in the spectral region, which can be measured with the same THz spectroscopy instrument on tailored samples.

%--------------------------------------------------------------------------------------------------

\begin{acknowledgements}
The authors thank R. Bittl (FU Berlin) and A. Schnegg (HZB) for permitting measurements at the FTIR-instrument, D. Ponwitz for technical support at the beamline, as well as Boris Eberhard, Peidong Yu, and Zach Evenson for the help with the measurements during the beamtime. Philip Born thanks Andreas Meyer for his continued support of the project. Financial support by the DFG research unit FOR1394 is gratefully acknowledged.
\end{acknowledgements}

%--------------------------------------------------------------------------------------------------

%--------------------------------------------------------------------------------------------------

\end{document}